\documentclass[aip,apl,reprint,twoside,floatfix,superscriptaddress]{revtex4-1}

\usepackage{graphicx}
\usepackage{amsmath}
\usepackage{amssymb}
\usepackage{amsfonts}
\usepackage{dcolumn}
\usepackage{epsfig}
\usepackage{subfigure}
\usepackage{bm}
\usepackage[version=3]{mhchem}
\usepackage{textcomp}
\usepackage{miller}
\usepackage{gensymb}

\begin{document}

\title{Percolation conductivity in hafnium sub-oxides}

\author{D.~R.~Islamov}\email{damir@isp.nsc.ru}
	\affiliation{Rzhanov Institute of Semiconductor Physics,
		Siberian Branch of Russian Academy of Sciences,
		Novosibirsk, 630090, Russian Federation}%
	\affiliation{Novosibirsk State University,
		Novosibirsk, 630090, Russian Federation}%
\author{V.~A.~Gritsenko}\email{grits@isp.nsc.ru}
	\affiliation{Rzhanov Institute of Semiconductor Physics,
		Siberian Branch of Russian Academy of Sciences,
		Novosibirsk, 630090, Russian Federation}%
	\affiliation{Novosibirsk State University,
		Novosibirsk, 630090, Russian Federation}%
\author{C.~H.~Cheng}
	\affiliation{Dept. of Mechatronic Technology, National Taiwan Normal University, Taipei, 106, Taiwan ROC}%
\author{A.~Chin}\email{albert\_achin@hotmail.com}
	\affiliation{National Chiao Tung University, Hsinchu, 300, Taiwan ROC}%

\date{\today}

\begin{abstract}
In this study, we demonstrated experimentally
that formation of chains and islands
of oxygen vacancies in hafnium sub-oxides (\ce{HfO_x}, $x<2$)
leads to percolation charge transport in such dielectrics.
Basing on the model of \'{E}fros-Shklovskii percolation
theory good quantitative agreement between the experimental
and theoretical data of current-voltage characteristics were achieved.
Based on the percolation theory suggested model shows that
hafnium sub-oxides consist of mixtures of
metallic Hf nanoscale clusters of $1\text{--}2\text{\,nm}$
distributed onto non-stoichiometric \ce{HfO_x}.
It was shown that reported approach 
might describe low resistance state current-voltage characteristics
of resistive memory elements based on \ce{HfO_x}.
\end{abstract}



\maketitle

Hafnium oxide (hafnia, \ce{HfO2}) and sub-oxides (\ce{HfO_x}, $x<2$)
play extremely important roles in modern microelectronics.
Hafnia is used in modern MOSFETs as high-$\kappa$ gate dielectric
with low leakage currents
\cite{Xu:JPL91:10127, Ma:JDMR5:36, Robertson:RPP69:327, Vandelli:ED58:2878}.
Hafnium sub-oxides are the most promising materials
to be used as active medium in 
resistive random access memory (RRAM)
\cite{Goux:APL97:243509, PhysRevB.85.195322},
which might be used for
universal memory combining  the most favorable
properties of both high-speed dynamic random access memory,
and non-volatile flash memory
\cite{Strukov:Nature453:80, rram:jjyang, doi:10.1021:nl803387q, Borghetti:Nature464:873, NatureMaterials:10:625}.
RRAM has many advantages:
 a simple metal-insulator-metal (MIM) structure,
 a small memory cell,
 potential for 3D integration,
 high read and write operation speeds,
 low power consumption,
 and the ability to store information over the long term.
A RRAM operation is principally based on switching back and forth
from the insulating medium's high resistance state (HRS)
to a low resistance state (LRS) when a current flows.
Conductivity of \ce{HfO_x} used as RRAM active medium
is limited by ionization of charge carrier traps
when RRAM is switched to the HRS state
\cite{PhysRevB.85.195322, PhysRevB.61.10361, HfO2:Tranport:2014}.
Unfortunately, unlike the flash memory,
the fundamental physics mechanism
of RRAM is still inadequate, but that is vital to realize
the ultra-low-power memory array.

In this letter, we report that the charge transport mechanism
is described in terms of percolation theory
when hafnium-sub-oxides-based RRAM is switched to the LRS state.

Transport measurements were recorded for MIM structures of
\ce{Si}/\ce{TaN}/\ce{HfO_x}/\ce{Ni}.
To fabricate these structures, we deposited the 8-nm-thick amorphous
\ce{HfO_x} on 100-nm-thick \ce{TaN} films
on \ce{Si} wafers, using physical vapor deposition.
A pure \ce{HfO2} target was bombarded by an electron beam,
and \ce{HfO2} was deposited on the wafer.
No post deposition annealing was applied to produce highly
non-stoichiometric \ce{HfO_x} films.
Structural analysis showed that the resulting 
films were amorphous.
All samples for transport measurements were equipped
with round 50-nm-thick \ce{Ni} gates with a radius of $70$\,$\mu$m.

Transport measurements were performed using
a Hewlett Packard 4155B Semiconductor Parameter Analyzer
and an Agilent E4980A Precision LCR Meter.
All measurement equipment
were protected against short circuiting
with the current through the sample limitation of $1$\,$\mu$A.

The most commonly used LRS description
in RRAM structures consists of conductive filament (CF),
approximately $1$--$10$\,nm in diameter ($D$)
\cite{Bersuker:JAP110:124518, Hou:APL98:103511, Ielmini:JEDL61:2378}.
The CF forming is cause by charged ion movements
due to temperate gradients and electric fields \cite{Ielmini:JEDL61:2378}.
It was supposed, that CFs in \ce{Ni}/\ce{HfO2}/\ce{Si}
RRAM structures consist of nickel, migrated from the metal
electrode \cite{Hou:APL98:103511},
but these assumptions were based on results of
RRAM measurements in 
\ce{Ni}/\ce{NiO}/\ce{Ni}-type structures (i.e. with \ce{NiO}
dielectric medium with \ce{Ni} electrodes).
However, CF has metallic 
temperature dependence of resistance
\begin{equation}
  (R-R_0)\propto (T-T_0)
  \label{e:R_T:Bersuker}
\end{equation}
(here $R_0$ is the CF resistance at temperature $T_0$),
with average resistivity in three order smaller than
resistivity of pure \ce{Hf} metal \cite{Bersuker:JAP110:124518}.

\begin{figure}
  \includegraphics[width=\columnwidth]{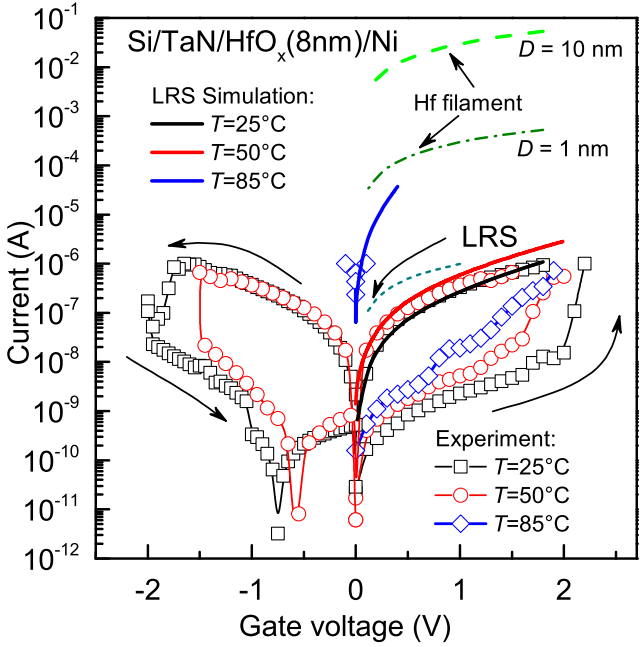}
  \caption{(Color online) Experimental RRAM current-voltage characteristics hysteresis (characters)
  in \ce{Si}/\ce{TaN}/\ce{HfO_x}/\ce{Ni} MIM structures at different temperatures.
  Black, red and blue solid lines present LRS simulations in terms of percolation model.
  Green lines model current-voltage characteristics of pure \ce{Hf} CF
  of length of $8$\,nm and diameter of $1$\,nm and $10$\,nm;
  dark cyan dotted line models $I$-$V$ of 
  sub-stoichiometric \ce{HfO_x} CF with diameter of $1$\,nm.}
  \label{f:IVT}
\end{figure}

Assuming, that the current-voltage characteristics ($I$-$V$)
are described by Ohm's law, we compared
calculated $I$-$V$ for Hf metal wire $8$\,nm long
with diameters of
$10$\,nm (dashed light green line in Fig.~\ref{f:IVT}),
$1$\,nm (dashed with dots dark green line in Fig.~\ref{f:IVT}),
and sub-stoichiometric \ce{HfO_x} CF \cite{Bersuker:JAP110:124518} 
($D=1$\,nm, dark cyan dotted line in Fig.~\ref{f:IVT})
with experiment results,
indicated by the colored characters in Fig.~\ref{f:IVT}.
Calculated values of the current through pure Hf wires
are $10^3$--$10^5$ times higher than the values
obtained in the experiments.
Furthermore, such a high current cannot be used for large memory
arrays because of high power consumption \cite{Cheng:ADMA201002946}.
We understand, that a cylinder shape of the metal filament
is not accurate model of CF, however we used these
estimations as motivations for further research.
The current through non-stoichiometric \ce{HfO_x} CF
is close to experiment results, but expected values
are large for temperatures of $T=25$--$50$\degree C,
and lower than measured at  $T=85$\degree C.
The current grows exponentially with temperature growing,
while following results form the literature,
the current should decrease with the temperature increasing
(\ref{e:R_T:Bersuker}).

Therefore, we suppose that LRS conductivity
is conditioned by the presence
of a non-stoichiometric \ce{HfO_x} islands in \ce{HfO2} matrix
as \ce{HfO_y} with $y\lesssim 1.89$ splits into phases of
\ce{Hf}, \ce{HfO_x} and \ce{HfO2}  \cite{HfOx:Fluctuations:2014}.
The overview on the structure of non-stoichiometric sub-oxides
and sub-nitrides was developed for \ce{SiO_x}, \ce{SiN_x},
and \ce{SiO_xN_y}
\cite{Hubner:JNCS36:1011, Novikov:JAP110:014107, Gritsenko:JNCS3297:96, PhysRevLett.81.1054}.

\begin{figure}
  \includegraphics[width=\columnwidth]{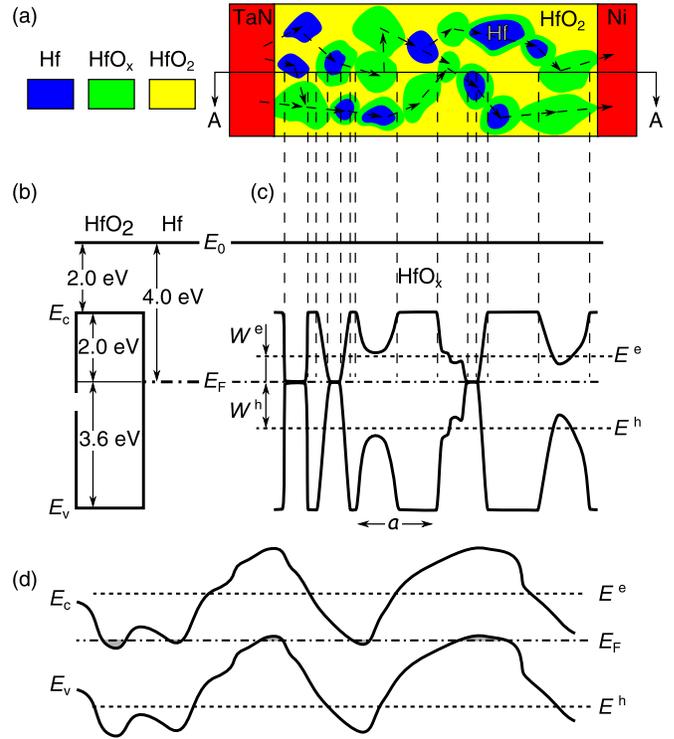}
  \subfigure{\label{f:PercModel:a}}
  \subfigure{\label{f:PercModel:b}}
  \subfigure{\label{f:PercModel:c}}
  \subfigure{\label{f:PercModel:d}}
  \caption{(Color online) Percolation model in electron systems with the nanoscale
  potential fluctuations.
  \subref{f:PercModel:a}~Schematic planar illustration of \ce{Hf}/\ce{HfO_x}
  ($x<2$) space-modulated by chemical composition structure.
  \subref{f:PercModel:b}~Flat band energy diagram of \ce{HfO2}/\ce{Hf} structure.
  \subref{f:PercModel:c}~Energy diagram of the structure with nanoscale potential fluctuations.
  \subref{f:PercModel:d}~Energy diagram of large-scale potential fluctuations in a semiconductor layer \cite{Shklovskii:en:1979}.
  $E^\textrm{e,h}$ and $W^\textrm{e,h}$ are percolation levels and percolation thresholds of the electrons and holes respectively.}
  \label{f:PercModel}
\end{figure}

A 2D structural image of non-stoichiometric \ce{HfO_x},
regarding the intermediate structural model \cite{Novikov:JAP110:014107},
is presented in Fig.~\ref{f:PercModel:a}.
According to this model, the CF in hafnium sub-oxide
consists of a mixture of
metallic hafnium nanoscale clusters (blue drops)
and non-stoichiometric \ce{HfO_x} (green islands) distributed
onto \ce{HfO2} matrix (yellow area).
Fig.~\ref{f:PercModel:c} is an energy diagram of \ce{HfO_x} in the intermediate structural model.
According to this plot, spatial fluctuations in the chemical composition
of \ce{HfO_x} lead to local band gap width spatial fluctuations.
The maximal fluctuation of the energy scale is equal to the \ce{HfO2}
band gap width of $E_\mathrm{g}=5.6$\,eV \cite{afanas2010internal}.
The work function of metallic hafnium is $4.0$\,eV.
The maximal fluctuation scale of the \ce{HfO_x} conduction band is $2.0$\,eV,
which equals the electron barrier height of \ce{Hf}/\ce{HfO2} interface.
The hole energy barrier of \ce{Hf}/\ce{HfO2} is $3.6$\,eV
(Fig.~\ref{f:PercModel:b}), which leads to 
the maximal fluctuation scale of the \ce{HfO_x} valence band of $3.6$\,eV.

The nanoscale fluctuations at the bottom of conduction band $E_\textrm{c}$ and
at the top of valence band $E_\textrm{v}$ are close to those proposed in the model developed in
\cite{Shklovskii-Efros:en:1975, Shklovskii:en:1979}, as shown in Fig.~\ref{f:PercModel:d}.
The charge transport in such electron systems can be described according to percolation.
This model assumes that excited electrons with energy higher than the flow level $E^\textrm{e}$ are delocalized,
and driving round a random potential, transfer the charge.
The hole conductivity is realized through the excitation of electrons
with energy $E^\textrm{h}$ to the Fermi level.
These excitations form hole-type quasiparticles,
which transfer the charge.
In other words, to be involved in transport processes,
electrons and holes must overcome energy thresholds ($W^\textrm{e,h}$ here, and $W^\textrm{e}\neq W^\textrm{h}$ in general).
The current-voltage characteristics are exponentials \cite{Shklovskii:en:1979}:
\begin{equation}
  I(T)=I_0(T)\exp\left( \frac{(CeFaV_0^\nu)^{\frac{1}{1+\nu}}}{kT} \right),
  \label{eq:percolation}
\end{equation}
where
$I$ is the current,
$I_0$ is the preexponential factor,
$e$ is the electron charge,
$F$ is the electric field,
$a$ is the space scale of fluctuations,
$V_0$ is the amplitude of energy fluctuation,
$k$ is the Boltzmann constant,
$C$ is a numeric constant,
and $\nu$ is a critical index.
The values of the constants were derived from Monte-Carlo simulations and evaluated at
$C\simeq 0.25$ \cite{Shklovskii:en:1979} and $\nu=0.9$ \cite{Shklovskii-Efros:en:1975}.
Percolation energy threshold $W$ can be evaluated based on the temperature dependency of
the preexponential factor:
\begin{equation}
  I_0(T)\sim\exp\left(-\frac{W}{kT}\right).
  \label{eq:I0}
\end{equation}

The solid colored lines in Fig.~\ref{f:IVT} indicate
the results of LRS simulations regarding the percolation model, given in (\ref{eq:percolation}).
Numeric fitting returns the value of combination as $CaV_0^{0.9}=0.45$\,$\text{nm}\cdot\text{eV}^{0.9}$,
which corresponds to $V_0=1.9$\,eV when $a=1$\,nm and $C=0.25$.
The slope of a fitting line in a $\ln(I_0)$-vs-$T^{-1}$ plate
according to (\ref{eq:I0}) corresponds to a percolation threshold
of $W\approx 1.0$\,eV.
Because $W\lesssim V_0 \leqslant 2.0\text{\,eV}$ (for electrons),
we can estimate the space size of nanoscale fluctuations as
$a\approx 1\text{--}2\text{\,nm}$.

Previous experiments in charge transfer have demonstrated that hafnia conductivity
is bipolar (or two-band) \cite{APL201108, Ando:EDL32:865, Vandelli:ED58:2878, NovikovHfO2:JAP113:024109}:
electrons are injected from a negatively shifted contact in the dielectric,
and holes are injected from a positively shifted electrode in the dielectric.
In our model, LRS conductivity can also be studied using electrons and holes.
For the reason of simplicity, the current study was limited
to considering monopolar electron conductivity.

The results demonstrate that charge transport in
non-stoichiometric hafnium sub-oxides 
is described according
to the percolation model in electron systems exhibiting potential
nanoscale fluctuations.
This approach can be applied to explain 
RRAM in \ce{GeO_x}- and \ce{SiO_x}-based  structures
\cite{Shaposhnikov:APL100:243506, SiO2RRAM:JAP111:074507}.

This work was partly supported by National Science Council, Taiwan
(grant No.~NSC103-2923-E-009-002-MY3)
(growing test structures, preparing samples, and performing transport measurements),
and by the Russian Science Foundation (grant No.~14-19-00192)
(calculations and modeling).

\bibliographystyle{apsrev4-1}
	\bibliography{IEEEabrv,../../../bibtex/Technique,../../../bibtex/GeO2,../../../bibtex/TiO2,../../../bibtex/La2O3,../../../bibtex/HfO2,../../../bibtex/SiO2,../../../bibtex/TaOx,../../../bibtex/ZrO2,../../../bibtex/Theory,../../../bibtex/percollation,../../../bibtex/Memristor,../../../bibtex/minecite}

\end{document}